\documentclass[conference, 10pt]{IEEEtran}
\IEEEoverridecommandlockouts

\usepackage[utf8]{inputenc}
\usepackage[stretch=50,shrink=50,kerning=true,spacing=true]{microtype}

\usepackage{amsmath,amssymb,bm}

\newcommand{\N}{\mathbb{N}}
\newcommand{\R}{\mathbb{R}}
\newcommand{\C}{\mathbb{C}}
\newcommand{\E}{\bm{\mathrm{E}}}

\DeclareMathOperator*{\Tr}{tr}
\DeclareMathOperator*{\Argmax}{argmax}

\usepackage[sorting=none,maxbibnames=3,
giveninits=true,style=ieee]{biblatex}

\usepackage{tikz}
\usetikzlibrary{positioning}
\usetikzlibrary{external}
\usepackage{pgfplots}
\DeclareUnicodeCharacter{2212}{−}
\usepgfplotslibrary{groupplots,dateplot}
\usetikzlibrary{patterns,shapes.arrows}
\pgfplotsset{compat=1.17}

\usepackage{siunitx}

\usepackage{xcolor}

\usepackage{soul}

\usepackage[hidelinks]{hyperref}
\usepackage{cleveref}
\usepackage[acronyms]{glossaries}
\newacronym{admm}{ADMM}{Alternating Directions of Multipliers Method}
\newacronym{anm}{ANM}{Atomic Norm Minimization}
\newacronym{adc}{ADC}{Analog-to-Digital Converter}
\newacronym{awgn}{AWGN}{Additive White Gaussian Noise}
\newacronym{asic}{ASIC}{Application Specific Integrated Circuit}
\newacronym{arpack}{ARPACK}{ARnoldi PACKage}
\newacronym{api}{API}{Application Programmable Interface}
\newacronym[plural=AOI, firstplural=Areas of Interest (AOI)]{aoi}{AOI}{Area of Interest}

\newacronym{bp}{BP}{Basis Pursuit}
\newacronym{bpdn}{BPDN}{Basis Pursuit Denoising}

\newacronym{crb}{CRB}{Cramér-Rao (lower) Bound}
\newacronym{cs}{CS}{Compressed Sensing}
\newacronym{cf}{CF}{Crest factor}
\newacronym{cr}{CR}{Compression Ratio}
\newacronym{cmos}{CMOS}{Complementary Metal Oxide Semiconductor}
\newacronym{cots}{COTS}{Commercial-Off-The-Shelf}
\newacronym{cpu}{CPU}{Central Processing Unit}
\newacronym{cvx}{CVX}{Convex Optimization toolboX}

\newacronym{doa}{DoA}{Direction of Arrival}
\newacronym{dod}{DoD}{Direction of Departure}
\newacronym{cM}{cM}{centiMorgan}
\newacronym[plural=DMC,firstplural=Dense Multipath Components (DMC)]{dmc}{DMC}{Dense Multipath Components}
\newacronym{das}{DAS}{Delay-and-Sum}
\newacronym{dft}{DFT}{Discrete Fourier Transform}
\newacronym{idft}{IDFT}{Inverse Discrete Fourier Transform}
\newacronym{dct}{DCT}{Discrete Cosine Transform}
\newacronym{dsp}{DSP}{Digital Signal Processor}

\newacronym{eadf}{EADF}{Effective Aperture Distribution Function}
\newacronym{esprit}{ESPRIT}{Estimation of Signal Parameters via Rotational Invariance Techniques}
\newacronym{ett}{ETT}{Eigenvalue Threshold Test}
\newacronym{eft}{EFT}{Exponential Fitting Test}

\newacronym{fht}{FHT}{Fast Hadamard Transform}
\newacronym[longplural={Fast Fourier Transforms}]{fft}{FFT}{Fast Fourier Transform}
\newacronym{fmcw}{FMCW}{Frequency-Modulated Continuous-Wave}
\newacronym{fpga}{FPGA}{Field Programmable Gate Array}
\newacronym{fri}{FRI}{Finite Rate of Innovation}
\newacronym{fim}{FIM}{Fisher Information Matrix}
\newacronym{fmc}{FMC}{Full Matrix Capture}
\newacronym{fista}{FISTA}{Fast Iterative Shrinkage-Thresholding Algorithm}
\newacronym{frvm}{FRVM}{Fast Relevance Vector Machine}

\newacronym{gfcs}{grid-free CS}{grid-free compressive sensing}
\newacronym{gpu}{GPU}{Graphical Processing Unit}
\newacronym{gan}{GAN}{Generative Adversarial Network}

\newacronym{hrpe}{HRPE}{High Resolution Parameter Estimation}

\newacronym{ir}{IR}{Impulse Response}
\newacronym{iid}{iid}{independent and identically distributed}
\newacronym{irf}{IRF}{Impulse Response Function}
\newacronym{ista}{ISTA}{Iterative Shrinkage-Thresholding Algorithm}

\newacronym{lasso}{LASSO}{Least Absolute Shrinkage and Selection Operator}
\newacronym{lse}{LSE}{Line Spectral Estimation}
\newacronym{lfsr}{LFSR}{Linear Feedback Shift Register}
\newacronym{lo}{LO}{Local Oscillator}
\newacronym{los}{LOS}{Line of Sight}
\newacronym{lam}{LAM}{Large Area Monitoring}

\newacronym{mimo}{MIMO}{multiple input multiple output}
\newacronym{mmv}{MMV}{multiple measurement vectors}
\newacronym{ml}{ML}{maximum likelihood}
\newacronym{mse}{MSE}{mean squared error}
\newacronym{bce}{BCE}{Binary Crossentropy}
\newacronym{mlbs}{MLBS}{Maximum Length Binary Sequence}
\newacronym{msm}{MSM}{M-Sequence Method}
\newacronym{mwc}{MWC}{Modulated Wideband Converter}
\newacronym{mpu}{MPU}{Microprocessor Unit}
\newacronym{mri}{MRI}{Magnetic resonance imaging}
\newacronym{music}{MUSIC}{Multiple Signal Classification}

\newacronym{ndt}{NDT}{Nondestructive Testing}
\newacronym{nde}{NDE}{Nondestructive Evaluation}
\newacronym{nn}{NN}{Neural Net}

\newacronym{omp}{OMP}{Orthogonal Matching Pursuit}
\newacronym{oop}{OOP}{Object Oriented Programming}

\newacronym{pdp}{PDP}{Power Delay Profile}
\newacronym{pn}{PN}{Pseudo-Noise}
\newacronym{pwc}{PWC}{Plane Wave Compounding}
\newacronym{pwi}{PWI}{Plane Wave Imaging}
\newacronym{pura}{PURA}{Patch Uniform Rectangular Array}
\newacronym{ura}{URA}{Uniform Rectangular Array}

\newacronym{rd}{RD}{Random Demodulator}
\newacronym{rem}{REM}{Reconstruction Error Metric}
\newacronym{rmse}{RMSE}{root mean squared error}
\newacronym{rms}{RMS}{root mean squared}
\newacronym{ric}{RIC}{Restricted Isometry Constant}
\newacronym{rip}{RIP}{Restricted Isometry Property}
\newacronym{rc}{RC}{Raised Cosine}
\newacronym{roi}{ROI}{Region of Interest}
\newacronym{rimax}{RIMAX}{Richter Maximization Approach}
\newacronym{rvm}{RVM}{Relevance Vector Machine}

\newacronym{scf}{SCF}{spatial correlation function}
\newacronym{sc}{SC}{Specular Components}
\newacronym{sdp}{SDP}{semi-definite program}
\newacronym{svd}{SVD}{singular value decomposition}
\newacronym{svm}{SVM}{Support Vector Machine}
\newacronym{soe}{SOE}{Sparsity Order Estimation}
\newacronym{sgd}{SGD}{Stochastic Gradient Descent}
\newacronym{stuca}{StUCA}{Stacked Uniform Circular Array}
\newacronym{spucpa}{SPUCPA}{Stacked Polarimetric Uniform Circular Patch Array}
\newacronym{suca}{SUCA}{Stacked Uniform Circular Array}
\newacronym{saft}{SAFT}{Synthetic Aperture Focusing Technique}
\newacronym{sota}{SOTA}{State of the Art}
\newacronym{ssd}{SSD}{Solid State Device}
\newacronym{ssr}{SSR}{Sparse Signal Recovery}
\newacronym{sa}{SA}{Synthetic Aperture}
\newacronym{spw}{SPW}{Single Plane Wave}
\newacronym{shm}{SHM}{Structural Health Monitoring}
\newacronym{snr}{SNR}{Signal-to-Noise Ratio}
\newacronym{stela}{STELA}{Soft-Thresholding with Exact Line Search Algorithm}
\newacronym{siso}{SISO}{Single Input Single Output}

\newacronym{th}{T\&H}{Track and Hold}
\newacronym{twista}{TWISTA}{Two-step Iterative Shrinkage-Thresholding Algorithm}

\newacronym{uca}{UCA}{uniform circular array}
\newacronym{ula}{ULA}{Uniform Linear Array}
\newacronym{uwb}{UWB}{Ultra-Wideband}
\newacronym{usndt}{US-NDT}{Ultrasonic Non-destructive Testing}

\makeglossaries

\AtEveryBibitem{
\clearlist{address}
\clearfield{month}
\clearfield{editor}
\clearfield{eprint}
\clearfield{volume}
\clearfield{isbn}
\clearfield{issn}
\clearfield{pages}
\clearlist{location}
\clearlist{pages}
\clearlist{editor}
}

\usepackage{placeins}

\begin{document}
\abovedisplayskip      = 2pt plus 1pt minus 1pt
\abovedisplayshortskip = 2pt plus 1pt minus 1pt
\belowdisplayskip      = 2pt plus 1pt minus 1pt
\belowdisplayshortskip = 2pt plus 1pt minus 1pt
\abovecaptionskip 		 = -5pt plus 1pt minus 1pt
\belowcaptionskip 		 = -3pt plus 1pt minus 1pt

\title{Estimating Multi-Modal Dense Multipath Components using Auto-Encoders}

\author{\IEEEauthorblockN{
S. Schieler\IEEEauthorrefmark{1},
M. Döbereiner\IEEEauthorrefmark{2},
S. Semper\IEEEauthorrefmark{1},
M. Landmann\IEEEauthorrefmark{2},
}

\IEEEauthorblockA{\IEEEauthorrefmark{1}
Electronic Measurements and Signal Processing Research Group,\\Technische Universit\"at Ilmenau, Ilmenau, Germany, steffen.schieler@tu-ilmenau.de
}

\IEEEauthorblockA{\IEEEauthorrefmark{2}
Fraunhofer Institute for Integrated Circuits, Ilmenau, Germany
}
\thanks{Supported by the Free State of Thuringia with funds from the European Social Fund.}
\thanks{S. Semper is funded by DFG under the project ``HoPaDyn'' Grant-No. TH 494/30-1.}
}


\maketitle

\begin{abstract}

We present a maximum-likelihood estimation algorithm for radio channel measurements exhibiting a mixture of independent \acrlong{dmc}.
The novelty of our approach is in the algorithms initialization using a deep learning architecture.
Currently, available approaches can only deal with scenarios where a single mode is present.
However, in measurements, two or more modes are often observed.
This much more challenging multi-modal setting bears two important questions: How many modes are there, and how can we estimate those?

To this end, we propose a \acrlong{nn}-architecture that can reliably estimate the number of modes present in the data and also provide an initial assessment of their shape.
These predictions are used to initialize for gradient- and model-based optimization algorithm to further refine the estimates.

We demonstrate numerically how the presented architecture performs on measurement data and analytically study its influence on the estimation of specular paths in a setting where the single-modal approach fails.

\end{abstract}

\IEEEpeerreviewmaketitle

\section{Introduction}\label{intro}
%
Radio channels, especially their accurate description, have been a vividly studied topic for the past decades.
Often, one seeks underlying parameters of the propagating waves in a measured environment.
Usually, the assumption is that these waves travel as specular rays/paths, i.e., \gls{sc}, through the environment in the form of plane waves.
However, it has been acknowledged~\cite{richter_estimation_2005,kaske_maximum-likelihood_2015,poutanen2011angledmc,vitucci2012dmc} that the the data model applied for \gls{hrpe} as a superposition of resolvable \gls{sc} is not sufficient to completely account for the bandlimited data collected by the receiver.
One possible physical interpretation of \glspl{dmc} is as a superposition of a large number of \gls{sc} which can not be resolved with bandwidth and \gls{snr} of the measurement system.

The authors in~\cite{landmann_impact_2012} demonstrate that the mentioned model mismatch can be overcome by accounting for this remaining energy as a colored Gaussian noise.
The underlying process has to be parametrized in a sophisticated manner such that it does capture the behavior of the non-resolvable components sufficiently well, does not interfere with the estimation of resolvable specular paths while also the complexity of the stochastic model allows for an efficient \gls{hrpe} process.

No matter if the process is modeled as both spatially and temporally correlated process~\cite{kaske_maximum-likelihood_2015}, or just temporally correlated~\cite{richter_estimation_2005,Oestges2012dmc}, the previously proposed models usually account for single mixture models, i.e., one \gls{dmc} mode, either in space- and delaytime or solely in delaytime.
However, for several propagation scenarios in different bands and bandwidths, see for instance~\cite{schneider2021measdata} and \Cref{measure}, we observed that the \gls{dmc} process must be modeled as a superposition of \emph{several} independent modes, which are separated in delaytime and/or space.
If this multitude of the \gls{dmc} is neglected during the estimation process, the wrongful estimation of a single mode ultimately also deteriorates the \gls{hrpe} of the \gls{sc} due to this inherent model mismatch and subsequently biased estimation of \gls{sc}.

In this paper, we focus on a multi-modal temporal distribution of the \gls{dmc} and assume a spatially uncorrelated process, for which \cite{richter_estimation_2005} presented a model-based maximum-likelihood approach, which is split in an initialization stage and an iterative refinement stage based on gradient descent. We give a summary of the algorithm in \Cref{sota_algo}. While the employed model itself can easily be extended to incorporate an arbitrary number of distinct modes, the initialization step implicitly assumes the presence of just one mode.

To allow for robust multi-modal estimation, we propose to use a so-called Auto-Encoder using convolutional networks~\cite{LeCun1989backprop}, which jointly infers the number and shapes of the \gls{dmc} modes directly from measured transfer-functions. Moreover, the proposed \gls{nn} can separate these, such that the initialization from \cite{richter_estimation_2005} can be applied separately. Along the way, we also propose a numerically more suitable parametrization of the \gls{dmc}.

\section{Algorithm}\label{algo}
We briefly review the setting for the currently available algorithms and explain their approach.
A radio channel observation consisting of $N_f \in \N$ frequency samples and $M$ uncorrelated snapshots can be modeled as
\begin{equation}\label{eqn:forward}
	\bm y = \bm f(\bm \theta) + \bm n(\bm \delta) + \bm n \in \C^{N_f \times M},
\end{equation}
where $\bm f : \Theta \subset \R^s \rightarrow \C^{N_f \times M}$ describes the resolvable specular components and $\bm n : \Delta \subset \R^d \rightarrow \C^{N_f \times M}$ is a complex zero-mean Gaussian but colored noise process we use to model the \gls{dmc}.
Additionally, the columns $\bm n_i$ of $\bm n$ are modeled as independent zero-mean Gaussian random vectors with covariance $\alpha_0 \bm I_{N_f}$ representing the measurement noise.

If we already have a sufficiently good $\hat{\bm \theta}$ to describe the specular components, it is justified to consider only the residual
\begin{equation}\label{eqn:dmc_forward}
	\bm r(\bm \theta) = \bm y - \bm f(\bm \theta) = \bm n(\bm \delta) + \bm n.
\end{equation}
Parametrizing $\bm n(\bm \delta) \sim \mathcal{N}(\bm 0_{N_f \times M}, \bm \Sigma(\bm \delta) \otimes \bm I_M)$, where $\otimes$ denotes the outer product, we see that the covariance of the zero-mean residual satisfies
\begin{equation}
	\E \left[\bm r({\bm \theta})^\ast \bm r({\bm\theta})\right]
		= (\bm \Sigma(\bm \delta) + \alpha_0 \bm I_{N_f}) \otimes \bm I_{M},
\end{equation}
where $\bm z^\ast$ denotes Hermitian transposition and $\bm \theta$ being the true parameter that is the origin of the observation $\bm y$.

\subsection{State of the Art Algorithm}\label{sota_algo}

\begin{figure}
\begin{center}
	\includegraphics[width=0.9\linewidth]{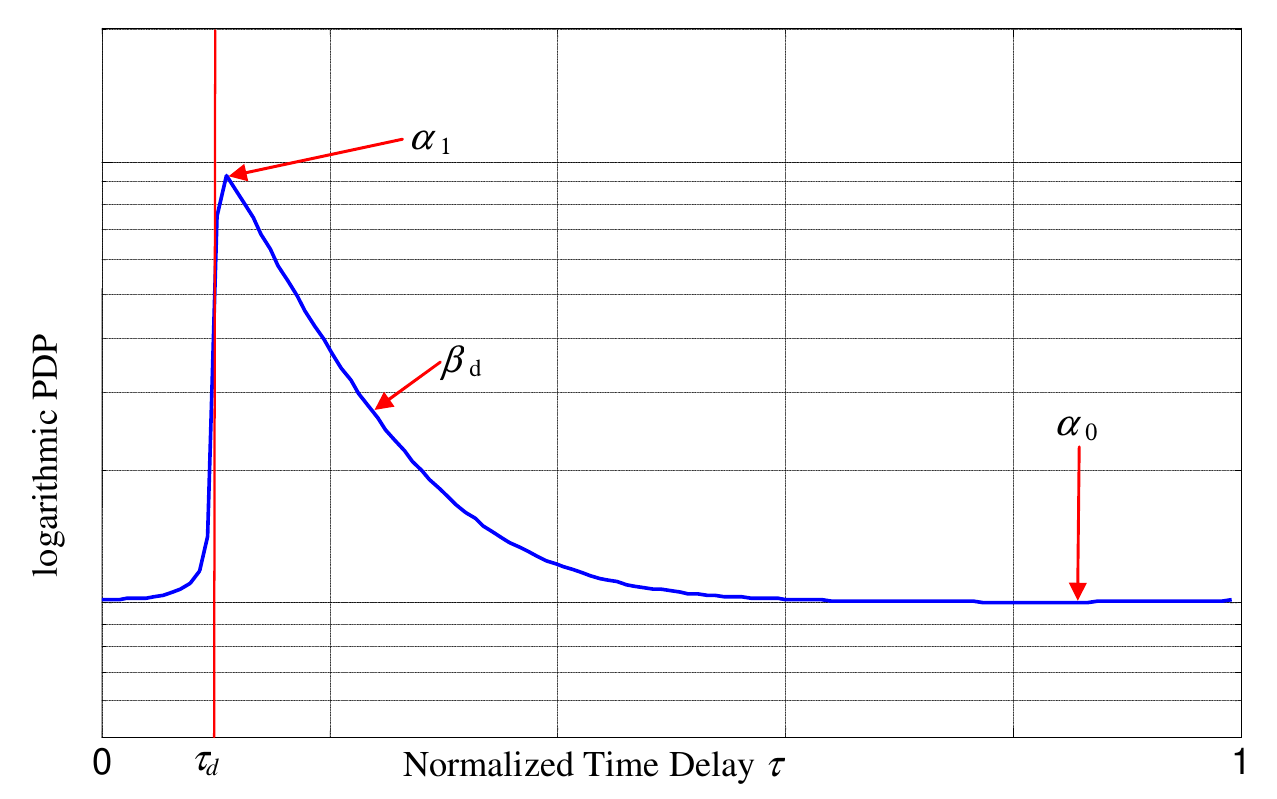}
\end{center}
\caption{Example for a single \gls{dmc} mode and its parameters\cite[p. 39]{richter_estimation_2005}. Here, we have that $\bm \delta = [\alpha_1, \beta_d, \tau_d]$.}\label{example_sigma}\vspace{-3mm}
\end{figure}

The core idea is to consider the negative log-likelihood of $\bm y$ with respect to $\bm \delta$ and $\bm \alpha_0$, which essentially reads as
\begin{equation}\label{llf_def}
	\lambda(\bm r, \bm \delta, \alpha_0) =
		\log \det (\bm \Sigma(\bm \delta) + \alpha_0 \bm I)
		+ \Tr (\bm r^\ast \bm \Sigma(\bm \delta)^{-1} \bm r),
\end{equation}
where we omitted constant summands, factors and the dependence on $\bm \theta$.
If we now consider \cite[eq. 2.61]{richter_estimation_2005} we find the mapping for $\bm \Sigma : \R^3 \rightarrow C^{N_f \times N_f}$ to be a Toeplitz matrix-valued function defined as
\begin{equation}\label{sigma_def}
	\bm \Sigma(\bm \delta)_{i,j} = \frac{
		\delta_1
	}{
		\delta_2 + \jmath 2 \pi (f_i - f_j)
	} \cdot \exp(
		- \jmath 2 \pi (f_i - f_j) \cdot \delta_3
	),
\end{equation}
where $f_i$ denotes the sampling frequencies used to obtain $\bm y$, where we renamed $\bm \delta = [\alpha_1, \beta_d, \tau_d]$.
See \Cref{example_sigma} for an interpretation of these.

\paragraph{Initialization}

In order to find a good estimate, one approach~\cite[Sec. 6.1.8]{richter_estimation_2005} is to first define $\hat{\bm r} = \sum_{i=1}^{M}\vert \bm F^\ast \bm r_i\vert^2$ using the Fourier matrix $\bm F$ and then to calculate
\[
	\hat{\alpha}_0 = \min \hat{\bm r};~~
	\hat{\delta}_1 = \max \hat{\bm r} - \hat{\alpha}_0
\]
together with
\[
	\hat{\delta}_2 = \frac{\hat{\delta}_1}{N_f(\Vert \hat{\bm r} \Vert_1 - \hat{\alpha}_0)},
\]
where $\Vert\cdot\Vert_1$ denotes the $\ell_1$-norm.
Finally, a simple yet usually sufficient estimate for $\delta_3$ is given by
\[
	\hat{\delta}_3 = \frac{\Argmax_i (\hat{\bm r}_{i+1} - \hat{\bm r_{i}}) - 1}{N_f-1}
\]
due to the shape of the assumed \gls{pdp}.

\paragraph{Refinement} Once a suitable $\hat{\bm \eta}^0 = [\hat{\bm \delta}^T, \hat\alpha_0]$ is found, we are free to perform any gradient-based algorithm.
For example, for our simulation and analysis we are going to use
\begin{equation}\label{lvm_updates}
	\bm \eta^{k+1} = \bm \eta^{k}
		+ (\bm H(\bm \eta^k) + \mu_k \bm I_4)^{-1} \cdot \bm J(\bm \eta^k),
\end{equation}
where $\bm H(\bm \eta^k)$ is the negative \gls{fim} defined via $\lambda$ and $\bm J(\bm \eta^k)$ is the score function of $\lambda$.
Simply plugging \eqref{sigma_def} into \eqref{llf_def} and using the definition for $\bm H$ and $\bm J$ allows to efficiently carry out the update step in \eqref{lvm_updates}.

\subsection{The Limitations}


If we consider the approach outlined above in the setting depicted in \Cref{crb_setup}, we see that the model mismatch in terms of number of \gls{dmc} modes is detrimental.
The only parameter correctly estimated is $\delta_3$, while the other account for the existence of two modes instead of the assumed single mode.
Not only is the estimate of $\bm \Sigma$ inherently biased, also any sensible estimation routine for the specular path parameters $\bm \theta$ is influenced by $\bm \Sigma$.
For instance, for specular paths arriving at a normalized delay in $[0.3, 0.5]$, the \gls{snr} is estimated to be worse than it actually is, which will have an influence on how reliable one considers the estimated parameter of this path.
In \Cref{influence} we provide a more detailed analysis of this phenomenon.

\subsection{Proposed Algorithm}

The modifications and extensions to the above state-of-the-art algorithm are three-fold.

\paragraph{Change of Variables}

Originally in \cite{richter_estimation_2005}, $\alpha_0$ and $\alpha_1$ were defined on a linear scale, which forced any optimization with respect to these parameters to obey the side constraints $\alpha_0,\alpha_1 > 0$.
However, these constraints can easily be alleviated by means of a change of variables as
\[
	\alpha^{\rm prev}_0
	= \exp(\alpha_0),\quad \alpha^{\rm prev}_1
	= \exp(\delta_1),
\]
which not only renders the optimization over $\bm \delta$ an unconstrained problem, but also the first- and second-order derivatives are much better behaved in terms of the condition number of the Hessian matrix.

\paragraph{Extension of the Model}
As a first step, we generalize the parametric model for $\bm \Sigma$ to the multi-modal version $\bm \Sigma_m : \R^{3 \times m} \rightarrow \C^{N_f \times N_f}$ via
\begin{equation}\label{sigma_generalized}
	\bm \Sigma_m(\bm \Delta) = \sum\limits_{i = 1}^m \bm \Sigma(\bm \Delta_i),
\end{equation}
which is a simple linear-combination of $m$ covariances corresponding to the single-mode setting.
This means implicitly that $\bm \Delta$ is the parameter of the multi-modal version of the \gls{dmc} process.
Hence, the quantities $\bm H$ and $\bm J$ used in \eqref{lvm_updates} can straightforwardly be extended to this more general model.
Consequently, the challenging task is robustly estimating the quantity $m$ as the number of modes present in $\bm y$ and their correct initialization.

\begin{figure}[t]
 \begin{center}
 	\input{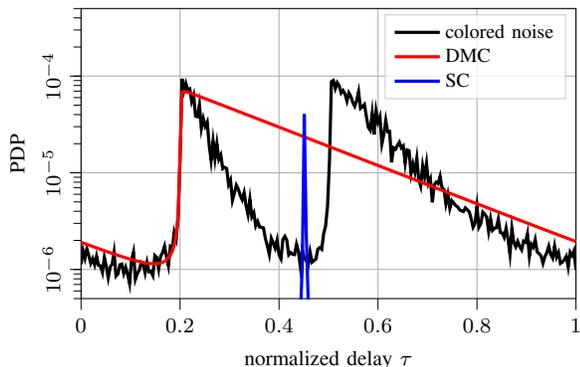}
 \end{center}
 \vspace{-10pt}
 \caption{Setup for the simulations of the \gls{crb} with two dense components. Initialization results for a single mode are shown in red. The simulated accuracy results are given in \Cref{crb_result}.}\label{crb_setup}\vspace{-3mm}
 \end{figure}

\paragraph{Initialization via an Auto-Encoder}
When looking at a \gls{pdp} like the one in \Cref{crb_setup}, it is intuitively clear how many modes are present and where they are located.
Hence, we approach this learning problem like a supervised-learning, $1$D imaging problem.
To this end, we designed an Auto-Encoder neural network targeted to solve two different tasks:
\begin{enumerate}
	\item Predict the correct model-order for the given data sample in the range $m=0,\dots,3$. Note the maximum value of \num{3} is arbitrary and the architecture can be extended to predict more modes straightforwardly.
	\item Recover, separate, and denoise up to $m=3$ \gls{dmc} modes present in $\bm y$, such that each component can be estimated separately with the already existing methods.
\end{enumerate}

With the denoised and separated modes obtained from the decoders, the actual parameters $\bm \delta$ of each mode are estimated using a Levenberg-Marquardt-algorithm based estimator.
In the following section, we outline the network architecture and the structure of the training data in order to accomplish these tasks.

\section{Learning Architecture}\label{net}

On a very high level, the architecture of the neural network consists of an encoder, which downsamples the complex-baseband input data into a latent space.
Then, up to three separate decoders are tasked with the reconstruction of the independent \gls{dmc} modes, based on the latent space.
\begin{figure}[t]
	\centering
	\resizebox{\columnwidth}{!}{
		\definecolor{enc}{HTML}{587BAC}
\definecolor{dec2}{HTML}{03C544}
\definecolor{dec1}{HTML}{1B8918}
\definecolor{dec0}{HTML}{014309}
\definecolor{mo}{HTML}{E27004}
\definecolor{lat}{HTML}{D7263D}
\tikzset{preprocess/.style={black,draw=black,fill=enc,rectangle,minimum height=1cm}}
\tikzset{encoder/.style={black,draw=black,fill=enc,rectangle,minimum height=2cm}}
\tikzset{decoder/.style={black,draw=black,rectangle,minimum height=2cm}}
\tikzset{latent/.style={black,draw=black,fill=enc,rectangle,minimum height=0.9cm}}
\tikzset{latentinner/.style={black,rectangle, minimum height=0.7cm, minimum width=0.7cm}}
\tikzset{modelorder/.style={black,draw=black,rectangle,minimum height=1cm}}

\begin{tikzpicture}
  \node[preprocess,rotate=90,minimum width=5.0cm, align=center] (preprocess) at (0,0) {\Large Preprocessing};

  \node[encoder,rotate=90,minimum width=5.0cm, align=center] (encoder) at (2,0) {\Large 4x\\\Large Conv + BN + ReLU};

  \node[latent,rotate=90,minimum width=2.3cm, label={[xshift=1.1cm, yshift=1.5cm]\Large Latent}, fill=lat] (latent) at (4,0) {};
  \node[latentinner, rotate=90, fill=dec0] (latentinner2) at (4,0.7) {};
  \node[latentinner, rotate=90, fill=dec1] (latentinner1) at (4,0) {};
  \node[latentinner, rotate=90, fill=dec2] (latentinner0) at (4,-0.7) {};

  \node[decoder, rotate=90, fill=dec0, minimum width=5.0cm] (decoder2) at (6.5+0.6, 0.3) {};
  \node[decoder, rotate=90, fill=dec1, minimum width=5.0cm] (decoder1) at (6.5+0.3, 0.0) {};
  \node[decoder, rotate=90, fill=dec2, minimum width=5.0cm, align=center] (decoder0) at (6.5, -0.3) {\Large 5x\\\Large Tr. Conv + BN + ReLU};

  \node[modelorder, rotate=90, fill=mo, minimum width=2.5cm, align=center] (modelorder0) at (6, -4.5) {\Large Conv + BN\\\Large + ReLU};
  \node[modelorder, rotate=90, fill=mo, minimum width=2.5cm, align=center] (modelorder1) at (7.5, -4.5) {\Large 2x\\\Large FC + ReLU};

  \node[align=left, minimum width=1cm, anchor=west, draw=dec0, ultra thick] (mode1) at (9, 1) {\Large Mode 1};
  \node[align=left, minimum width=1cm, anchor=west, draw=dec1, ultra thick] (mode2) at (9, 0) {\Large Mode 2};
  \node[align=left, minimum width=1cm, anchor=west, draw=dec2, ultra thick] (mode3) at (9, -1) {\Large Mode 3};
  \node[align=left, minimum width=1cm, anchor=west] (molabel) at (9, -4.5) {\Large Modelorder};

  \draw[->, ultra thick] (preprocess) -- (encoder);
  \draw[->, ultra thick] (encoder) -- (latent);
  \draw[->, ultra thick] ([xshift=0.2cm]latentinner2.south) -- ([xshift=1cm, yshift=1cm]latentinner2.south);
  \draw[->, ultra thick] ([xshift=0.2cm]latentinner1.south) -- ([xshift=1cm, yshift=0cm]latentinner1.south);
  \draw[->, ultra thick] ([xshift=0.2cm]latentinner0.south) -- ([xshift=1cm, yshift=-1cm]latentinner0.south);
  \draw[->, ultra thick] ([xshift=-.8cm]mode1.west) -- ([xshift=-.1cm]mode1.west);
  \draw[->, ultra thick] ([xshift=-.8cm]mode2.west) -- ([xshift=-.1cm]mode2.west);
  \draw[->, ultra thick] ([xshift=-.8cm]mode3.west) -- ([xshift=-.1cm]mode3.west);

  \draw[->, ultra thick] (latent.west) |- (modelorder0.north);
  \draw[->, ultra thick] (modelorder0.south) -- (modelorder1.north);
  \draw[->, ultra thick] (modelorder1.south) -- (molabel.west);

  \node[align=center, minimum width=2cm] (skip) at ([xshift=-.25cm,yshift=2.5cm]latent.east) {\Large Skip Connections};
  \draw[ultra thick] ([yshift=.1cm]encoder.east) -- ([xshift=.6cm, yshift=.9cm]encoder.east);
  \draw[ultra thick] ([xshift=.6cm, yshift=.9cm]encoder.east) -- ([xshift=3.0cm, yshift=.9cm]encoder.east);
  \draw[->, ultra thick] ([xshift=3.0cm, yshift=.9cm]encoder.east) -- ([xshift=3.6cm, yshift=.1cm]encoder.east);
\end{tikzpicture}
	}
	\vspace{.5mm}
	\caption{The architecture of the neural network is a 1D adaptation of U-Net. Additional decoders are added to predict multiple \gls{dmc} modes and a model-order subnet (orange) to predict the input model order based on the latent space also used for the reconstruction task.}
	\label{fig:neuralnetarchitecture}\vspace{-2mm}
\end{figure}
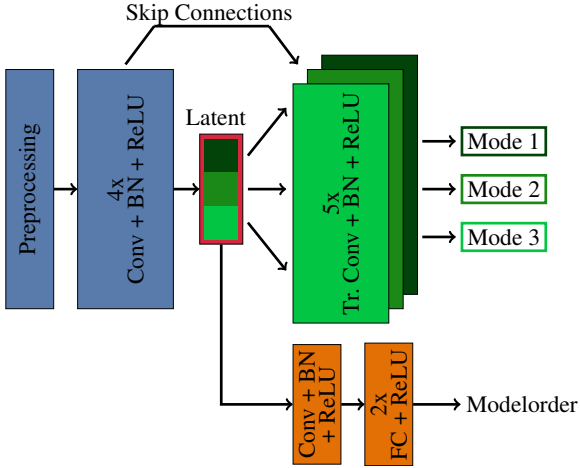
\paragraph{Labels and Data}

As we consider using real measurement data to train our algorithm infeasible, because it is not practical to manually generate a critical amount of labeled data, we decided to use a synthetic dataset for the training by means of the model $\bm \Sigma$.
This bears the advantage that we can generate an almost arbitrarily large dataset to train the network to obtain a well generalizing estimator.

Hence, our synthetic dataset consists of $10^8$ randomly generated instances of $\bm\Sigma_m(\bm \Delta) + \alpha_0 \bm I_{N_f}$ and the corresponding random realizations $\bm r = \bm n(\bm \Delta) + \bm n \in \C^{N_f \times M}$ as input data, ensuring the network never actually sees the exact same sample twice, to strengthen generalization and robustness to noise.

\begin{figure}
	\centering
	\input{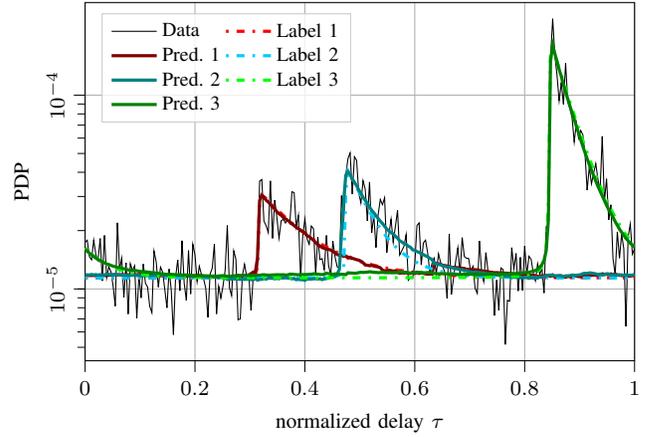}
	\caption{Data, labels and predictions of the proposed learning algorithm applied to synthetic data. The predictions are processed individually afterwards to estimate $\bm \Delta$. Periodicity is an artifact of the \gls{dft}-based preprocessing.}
	\label{fig:datastructure}\vspace{-2mm}
\end{figure}

As labels, we use the corresponding, separated \gls{dmc} components denoted by $\bm \Sigma(\bm \delta_m)$.
The relationship between input data, labels and the resulting predictions for the supervised learning task is illustrated by \Cref{fig:datastructure}.
\paragraph{Preprocessing}
Before being passed to our neural network, the input data $\bm r$ are preprocessed.
The preprocessing involves applying an inverse \gls{dft} along the frequency  dimension and using the mean of the magnitude-square
\begin{equation}\label{eqn:preprocess}
	\bm d = \frac{\sqrt{N_f}}{M} \sum_{k=1}^{M}
		\left \vert
			\bm F^\ast \bm r_k
		\right \vert^2 \in \C^{N_f},
\end{equation}
where $\bm F^\ast \in \C^{N_f \times N_f}$ denotes the inverse \gls{dft} matrix and $\bm r_k$ the $k-$th snapshot, where the contribution of \glspl{sc} has already been removed.

The second preprocessing step regards appropriate normalization of the input data $\bm d$ to limit the dynamic range of the input values fed into the non-linear neural network.
This step is necessary, as dynamic ranges in real-world measurement data can vary over several orders of magnitude.
To address this, we normalize the values in $\bm d$ to get the normalized version $\bm d_n$ by means of applying
\begin{equation}\label{sigma_generalized_normalized}
	\bm d_n = \log \bm d - \log \max (\bm d)
\end{equation}
element-wise.

This normalization scheme is calculated based on the inputs only and enables reconstruction of correctly scaled predictions after forward propagation using stored values of $\max (\bm d)$ and rescaling the forward propagation output accordingly.

\paragraph{Network Architecture}
The general design of the applied neural network architecture is lent from U-Net \cite{ronneberger_u-net_2015}.
As our task is based on $1$-dimensional input data, the $2$D convolutional layers in U-Net are replaced by appropriate $1$-D convolutions.
Before being passed to the encoder, $\bm d_n$ is passed through two convolutional layers (with batchnorm and ReLU activation function), which increases the number of channels from 1 to 32.
In the encoder four downsampling blocks are used, each consisting of a convolutional layer followed by batch normalization and ReLU activation function.
The number of filters is doubled by the convolution layers in each downsampling block, which are parametrized with a kernel size of \num{3}, stride of \num{2} (to achieve the downsampling), and circular padding of \num{1}.

After the downsampling, the latent space has a dimension of \num{16} features (with \num{512} channels).
A convolutional layer is used to upscale the latent to \num{24} features, enabling passing \num{8} features to each of the three decoders.
The decoder consists of three individual, structural identical, decoders (as we attempt to reconstruct $M \leq 3$ modes from each $\bm d_n$), each with 5 upsampling blocks.
Each upsampling block consists of a transposed convolution, followed by two convolutional layers.
Skip Connections between the decoder and encoders are added to the all but the first upsampling block after each transposed convolution layer.
Every decoder receives a third of the latent space to reconstruct its respective target mode.

The model order is prediction is performed by a small subnet attached to the latent space.
A convolutional layer is followed by three fully-connected layers to predict a one-hot encoded integer number between \numrange{0}{3}.

\paragraph{Loss and Training}
To ensure proper convergence of the neural network weights the used loss function is a weighted sum of two different components.
\begin{enumerate}
	\item the mode reconstruction loss uses \gls{mse}
	\item the modelorder loss uses \gls{bce}
\end{enumerate}

 For the computation of the mode reconstruction loss, the predicted model-order $\widetilde{m}$ is taken into account to mask mode predictions of non-existent modes in the decoders.
 For example, if the predicted model-order is \num{2}, the mode prediction of the third decoder is excluded from the computation of the MSE loss.
 This results in the following loss function $l :  \R^{N_f \times 3} \times \R^{N_f \times 3} \times \N \times \N \rightarrow \R^+_0$ defined as
 \begin{align}\label{eqn:loss}
	 l(\bm x, \tilde{\bm x}, m, \widetilde{m}) =& w_x \sum_{k=0}^{\widetilde{m}} \Vert \bm x_k -\tilde{\bm x}_k \Vert_2^2 \\ \nonumber
	 +& w_m \left[\widetilde{m} \log(m) + (1-\widetilde{m}) \log (1-m)\right]
 \end{align}
 where $\bm x$, $\tilde{\bm x}$ denote the label and prediction for the modes from the decoders, $m$ and $\tilde{m}$ denote the label and prediction for the model-order, and $w_x$ and $w_m$ denote scaling weights for the two different prediction parts.
 The loss weights were fixed to $w_x = 1$, $w_m = 100$, based on observations during training, such that both parts have similar magnitude.

\section{Simulations}\label{simu}
%
\subsection{Influence of the DMC on the Estimation Accuracy}\label{influence}
To showcase the importance of the correct \gls{dmc} model order for the estimation process, we want to study the resulting best-case accuracy any unbiased estimator can achieve by means of the \gls{crb}.
To this end, we create a simple synthetic setup, where the simulated \gls{ir} contains a single specular component with fixed path weights and fixed normalized delay $\tau=0.45$ is enclosed by two \gls{dmc} modes as depicted in \Cref{crb_setup}.
The analysis of the \gls{crb} is not only beneficial for theoretic purposes.
The estimator proposed in \cite{richter_estimation_2005} also uses the inverse \gls{fim} to determine which specular paths are reasonable estimates and possibly removes specific specular paths if the expected estimation variance indicated by the \gls{crb} is too high.

We simulate two different cases.
First, the estimation of a single mode, where one mode is spanning the area of the two modes, to account for both at the same time.
Second, we carry out the proposed method, which should correctly detect the two modes.
We run this scenario for varying intensity of the second mode by means of varying $\delta_1$ of this \gls{dmc}, i.e. $\Delta_{2,1}$.
Then, we evaluate the deterministic \gls{crb} for a peak at the depicted position, and we average these values over $200$ realizations per level of $\delta_1$ of the second mode.

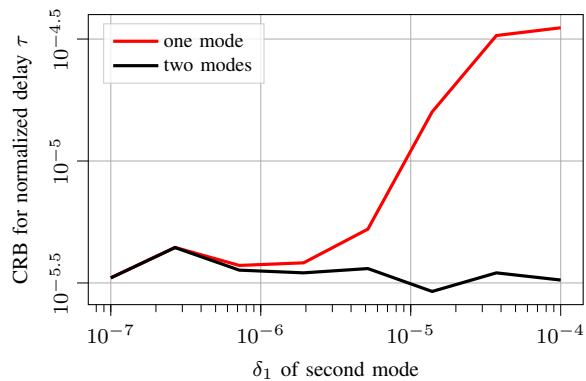
\begin{figure}[t]
\begin{center}
\begin{tikzpicture}

\begin{axis}[
colorbar style={yticklabel style={font=\footnotesize}, width=2mm},
height=0.3\textwidth,
label style={font=\footnotesize},
legend cell align={left},
legend style={
  fill opacity=0.8,
  draw opacity=1,
  text opacity=1,
  at={(0.03,0.97)},
  anchor=north west,
  draw=white!80!black
},
legend style={nodes={scale=0.75, transform shape}},
log basis x={10},
log basis y={10},
tick align=outside,
tick pos=left,
width=0.45\textwidth,
x grid style={white!69.0196078431373!black},
xlabel={\(\displaystyle \delta_1\) of second mode},
xmajorgrids,
xmin=7.07945784384137e-08, xmax=0.000141253754462275,
xmode=log,
xtick style={color=black},
xticklabel style={font=\footnotesize},
y grid style={white!69.0196078431373!black},
ylabel={CRB for normalized delay \(\displaystyle \tau\)},
ymajorgrids,
ymin=2.57716675548741e-06, ymax=3.98441768342093e-05,
ymode=log,
ytick style={color=black},
yticklabel style={rotate=90,font=\footnotesize}
]
\addplot [very thick, red]
table {%
1e-07 3.32085684084674e-06
2.68269579527973e-07 4.42179298628591e-06
7.19685673001151e-07 3.73128158331268e-06
1.93069772888325e-06 3.8246384071326e-06
5.17947467923121e-06 5.25597728087152e-06
1.38949549437314e-05 1.59013318372622e-05
3.72759372031494e-05 3.26946187631262e-05
0.0001 3.51810905810769e-05
};
\addlegendentry{one mode}
\addplot [very thick, black]
table {%
1e-07 3.31635824305724e-06
2.68269579527973e-07 4.41722317642593e-06
7.19685673001151e-07 3.5692511615882e-06
1.93069772888325e-06 3.47924590481023e-06
5.17947467923121e-06 3.62250428289218e-06
1.38949549437314e-05 2.91875795323177e-06
3.72759372031494e-05 3.47904638929581e-06
0.0001 3.25109136052767e-06
};
\addlegendentry{two modes}
\end{axis}

\end{tikzpicture}
\end{center}
\caption{\gls{crb} of the delay of specular component shown in \Cref{crb_setup} (blue) for one and two \gls{dmc} modes. If the estimator assumes only one mode present (red), increasing the energy in the second mode ($\delta_1$), degrades the estimators accuracy. Using two modes (black), the energy of the second mode can be increased without noteworthy changes in accuracy.}\label{crb_result}\vspace{-3mm}
\end{figure}

As we can see in \Cref{crb_result}, the resulting estimation accuracy differs significantly for the two settings.
For higher intensities of $\delta_1$, the scenario increasingly resembles the one depicted in \Cref{crb_setup}.
Hence, the assumed colored noise distribution renders the specular path harder to distinguish from random fluctuations in the measurement.
However, if we correctly impose a model with two modes, the intensity of the modes does not significantly influence the predicted accuracy for the specular paths delay.

\subsection{Joint Estimation of DMC and SC}

To also apply the proposed \gls{dmc} estimation routine to a more realistic setup, we used measurement data collected in a channel-sounding campaign, which clearly shows at least two \gls{dmc} modes in some of the snapshots~\cite{schneider2021measdata}.
However, the data still contains specular components that need to be removed from the observation, such that the assumption in \eqref{eqn:dmc_forward} is valid enough such that the \gls{pdp} is not biased by the remaining specular components.
Note, in practice the estimation of both the \gls{sc} and the \gls{dmc} is heavily intertwined, since the estimates of $\bm \theta$ and $\bm \delta$ are influencing each other.
This effect is also visible in the presented data. The mode starting at normalized delay $\tau \approx 0.7$ might not be estimated, if more specular components had been accounted for during the estimation prioir to the \glspl{dmc} estimation.

\begin{figure}[t]
\begin{center}
	\input{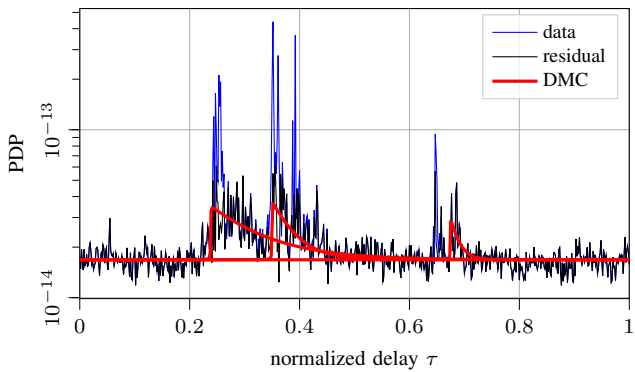}
\end{center}
\caption{Estimation of three \gls{dmc} modes in measurement data~\cite{schneider2021measdata} from a channel sounding campaign. The \gls{dmc} are estimated from the residual using the proposed neural network previously trained on synthetic data.}\label{measure}\vspace{-3mm}
\end{figure}

In \Cref{measure} we plot the acquired data $\bm y$, the residual $\bm r(\bm \theta)$ and $3$ estimated \gls{dmc} modes.
We used an algorithm similar to the one presented in~\cite{richter_estimation_2005} to get an estimate for $\bm \theta$ and hence $\bm r(\bm \theta)$.
To estimate $\bm \delta$, we used the model-order selection provided by the proposed \gls{nn} and then the initialization and refinement explained in \Cref{sota_algo}.
The first thing we notice is that the prediction and estimation work on measurement data, which is not to be taken for granted since the architecture was trained on synthetic data without any specular components.
Second, the results indicate that the proposed architecture is also able to estimate overlapping modes quite robustly.
However, depending on the use-case of the estimation data, it could be debatable if the third mode is necessary to be estimated.

\section{Conclusion}\label{concl}
Proper estimation of multiple \glspl{dmc} is required to obtain an unbiased estimate of both the dense as well as the specular components.
We show that the introduced autoencoder-based neural net design allows estimating the separated \glspl{dmc} modes together with model-order.
We show that although the network is trained on an artificially generated, synthetic dataset, the results suggest that it can also be applied to measurement data.
Our results show, the model-based neural network approach is suitable to denoise and separate up to three \glspl{dmc} and suitable for subsequent estimator initialization, surpassing other state-of-the-art approaches.

To further extend the current approach, two different directions seem the most useful for the problem at hand, i.e., direct estimation of \gls{dmc} parameters and extension to the angular domain.
First,
including the angular domain in the estimates provides a valuable framework extension for use cases in antenna array applications.
With additional angular information of the \glspl{dmc} better separation between dense components can likely be achieved as the individual \glspl{dmc} can also be separated with regard to their respective directivity angle in the array.
Second, to further improve the network performance on measurement data, it is also possible to tune the parameter distribution in the training set to better match the characteristics of the specific measurement dataset.
This could be accomplished by utilizing a \gls{gan} to generate training data which is even more similar to the measurement data encountered in the inference task.

\printbibliography

\end{document}